\documentstyle[12pt]{article}
\renewcommand{\baselinestretch}{1.5}
\begin{document}

\title{Low-lying S-wave and P-wave Dibaryons in a Nodal Structure Analysis}

\author{{Yu-xin Liu$^{1,3,4}$, Jing-sheng Li$^{1}$, and Cheng-guang Bao$^{2,4}$}\\
\noalign{\vskip 5mm}
 \normalsize{$^{1}$ Department of Physics, Peking University,
Beijing 100871, China} \\
\normalsize{$^{2}$ Department of Physics, Zhongshan University,
Guangzhou 510275, China} \\
\normalsize{$^{3}$ Institute of Theoretical Physics,
Academia Sinica, Beijing 100080, China} \\
\normalsize{$^{4}$ Center of Theoretical Nuclear Physics, National
Laboratory of}\\
\normalsize{Heavy Ion Accelerator, Lanzhou 730000, China} }

%\date{\today}

\maketitle

\begin{abstract}
The inherent nodal surface structure analysis approach is proposed
for the six-quark clusters with $u$, $d$ and $s$ quarks. The
wave-functions of the six-quark clusters are classified, and the
contribution of the hidden-color channels are discussed. The
quantum numbers and configurations of the wave-functions of the
low-lying dibaryons are obtained. The states
$[\Omega\Omega]_{(0,0^{+})}$, $[\Omega\Omega]_{(0,2^{-})}$,
$[\Xi^{*}\Omega]_{(1/2,0^{+})}$,
$[\Sigma^{*}\Sigma^{*}]_{(0,4^{-})}$ and the hidden-color channel
ones with the same quantum numbers are proposed to be the
candidates of experimentally observable dibaryons.
\end{abstract}

%{\bf Keywords:} dibaryon \quad cluster model\quad symmetry\quad
% inherent nodal surface  \quad QCD

%\end{keyword}

\section{Introduction} \label{sec:Intro}
It has been known that quark degrees of freedom are  essential to
describe properties of hadrons and the quantum chromodynamics(QCD)
is the underlying theory for strong interaction. One of the
current issues in this field involves the attempt to derive the
short-range part of the nucleon-nucleon interaction from the
quark-gluon degrees of freedom. Unfortunately, it is now very
difficult to make quantitative predictions with QCD at low and
intermediate energy region. Therefore various phenomenological
models based on the QCD assumptions, such as bag
models\cite{SPWcj}, cluster model\cite{SPWka} have been developed.

With the MIT bag model, a dibaryon, namely the H particle, was
predicted by Jaffe in the later of 1970's\cite{SPWja}. Thereafter
many attempts have been made to search for dibaryons in both
theory and experiment, because it is an appropriate place to
investigate the quark behavior in short distance and to explore
the exotic states of QCD. On the theoretical side, almost all QCD
inspired models (see for example
Refs.\cite{Muld80,Kond87,Gold87,Wang95,Kope908,ZY00}), even the
QCD sum rules\cite{SPWsumrule} and lattice QCD\cite{SPWlattice}
predict that there exist dibaryon states. However, no dibaryon
state except for the deuteron has been well established in
experiment up to now. On the other hand, the theoretical
predictions are quite far away from each other, since different
approaches take different models and interactions (at least the
interaction strength).

After Jaffe predicted the existence of the H particle(strangeness
$s= -2$, $J^{P}=0^{+})$ as a color-singlet multi-quark
system($q^{m}\overline{q}^{n}, n+m > 3$), Harvey developed a
method in cluster model\cite{SPWharvey}, with which all the
antisymmetric six-quark states with definite orbital and
isospin-spin symmetries were classified. Meanwhile, many
hidden-color channel states which can not be represented in terms
of free baryons were found. Unfortunately, the constituents Harvey
considered include only $u$ and $d$ quarks. In other words, the
strangeness of the system has not yet been taken into account.
However, more and more evidences proved that the system with
non-zero strangeness may be more stable to exist as a dibaryon(see
for example Refs.\cite{Gold87,Kope908,ZY00,LS00,SPW21}). It is
then necessary to extend Harvey's framework to include $s$ quarks
and get a complete result.

Along the way of the non-relativistic QCD, the concept of
wave-function can be implemented to describe many-quark states,
and embodied in the non-relativistic quark
model\cite{SPWharvey,SPWsuz}. On the other hand, it has been known
that analyzing the symmetric property and the inherent nodal
surface\cite{SPWbao1,SPWbao2,SPWbl} is a quite powerful approach
to investigate few-body electron and nucleon systems. To neglect
handling the complicated strong interactions of the six-quark
system in QCD and obtain model independent result, we will then
study the dibaryon states by extending Harvey's algebraic
framework and analyzing the inherent nodal surface of the system
in this paper.

This paper is organized as follows. Following this introduction,
the symmetry of the six-quark cluster which may include $u$, $d$
and $s$ quarks will be analyzed in Section 2. In Section 3, the
quantum numbers of the possible six-quark states will be discussed
systematically by analyzing the inherent nodal surface structure
of the system. In Section 4 and Section 5, we discuss the
low-lying S-wave and P-wave states and the perspective to search
for dibaryons, respectively. Finally, a summary and some remarks
are given in Section 6.

\section{Symmetries of the Six-quark States}\label{Class}

The wave-function of six-quark systems can be written as the
coupling of an orbital part and an internal part, fulfilling the
requirement that the total wave-function must be antisymmetric,
i.e. has a symmetry [$1^{6}$]. It has been well known that the
system with $u$, $d$ and $s$ quarks possesses the internal
symmetry
SU$_{CFS}(18)\supset$SU$_{C}$(3)$\otimes$SU$_{F}$(3)$\otimes$SU$_{S}$(2).
Let $[f]_{O}$, $[f]_{C}$ and $[f]_{FS}$ be the irreducible
representations (irreps) of the groups associated with the
orbital, color and flavor-spin space, respectively, we shall have
\begin{equation}
\label{g1}[1^{6}]\in[f]_{O}\otimes[f]_{C}\otimes[f]_{FS}.
\end{equation}
Although six-particle system in general does not impose any
constraint on the orbital wave-function, which may have any one of
the symmetries listed in the first column of Table 1, we should
determine which choices are more favorable to binding and
therefore more probable to exist as low-lying states.

The irrep $[f]_{CFS}$ of the SU(18) can be reduced to $[f]_{C}$ of
SU$_{C}$(3) and $[f]_{FS}$ of SU$_{FS}$(6) with the standard
method\cite{SPWch}. The lack of direct experimental observation of
free ``color charge" suggests that not only the free baryons but
also all the other observable states should be SU(3) color
singlet. We can thus restrict our study of six-quark systems to
those having color symmetry
\begin{equation}
\label{g2}[f]_{C}=[2\, 2 \, 2].
\end{equation}
Then we get the possible irreps $[f]_{FS}$ as listed in Table 1.

It has been known that $u$, $d$ and  $s$ quarks can construct
eight free baryons: N, $\Delta$, $\Sigma$, $\Sigma^{*}$,
$\Lambda$, $\Xi$, $\Xi^{*}$ and $\Omega$. All of them are the
building blocks of the dibaryons and have flavor-spin symmetry
$[f]_{FS}$=[3]. Thus it is reasonable to assume that the low-lying
dibaryon states which can be observed in experiments may have
symmetry
\begin{equation}
\label{g3}[f]_{FS}\in[3]\otimes[3]=\{ [6], [5\, 1], [4\, 2],
 [3\, 3]\}.
\end{equation}
 These particular
symmetries are shown by an asterisk in Table 1 for latter
reference.

According to the subclassification of the symmetry
SU$_{FS}\supset$U$_{s}$(1)$\otimes$SU$_{T}$(2)$\otimes$SU$_{S}$(2),
we list the strangeness, isospin and spin for the six-quark
systems in Table 2, where both the two-baryon bound states and the
hidden-color channel states (marked with CC) are listed. Table 2
shows that there exist many hidden-color channel states in the
six-quark systems, whose SU$_{FS}$(6) symmetry are the same as
those of the two-baryon states. Furthermore, all the states of the
SU$_{FS}$(6) symmetries without asterisks in Table 1 are
also``hidden-color" states in the cluster representation.

Furthermore, as a constituent of the low-lying dibaryons, each
baryon is assumed to be in its ground state. It turns out that all
these ground states have orbital symmetry $[f]_{O}=[3]$. It
follows that in the cluster model, we can consider the six-quark
system whose orbital wave-functions hold the symmetry
\begin{equation}
\label{g4}[f]_{O}\in[3]\otimes[3]= \{ [6], [5\, 1], [4\, 2],
 [3\, 3]\},
\end{equation}
if only the two-baryon bound states are taken into account. If we
include the hidden-color channel states, the other orbital
symmetries should consequently be deliberated. The realistic
configuration of the orbital wave-functions should be fixed with
the intrinsic property of the system.

\section{Nodal Structure Analysis of Low-lying Six-quark States}\label{sec:Symm}

The Schr\"{o}dinger equation of multi-particle system can be
written as:
\begin{equation}
\label{eq1} H\psi(\xi)=E\psi(\xi) \, ,
\end{equation}
where $\xi$ is the set of variables, such as coordinates, spins
and isospins. If $G$ is the symmetry group of the Hamiltonian $H$,
in other words, $H$ is invariant to the transformation $\hat{O}$
which is an element of the group $G$, the eigenstates of the
Hamiltonian can be classified with the irreducible representation
of $G$.

On the other hand, It has been known that if a state contains
excited spatial oscillations (i.e., the orbital wave function
contains nodal surfaces if observed in a body frame), it would be
higher in energy than the states not containing excited
oscillations. For example, the energy and the number $n$ of nodals
of the states in one-dimensional infinite wall is $E_{n} =
\frac{\pi ^2 \hbar ^2 (n + 1)^2}{2 m a^2}$, where $a$ is the width
of the wall and $m$ is the mass of the particle. Moreover, It has
been found that there are two kinds of nodal surfaces. The first
kind of them depends on dynamics, while the second relies purely
on symmetry. It means that the second kind nodal surfaces are
inherently contained in certain wave functions. Let $\Psi $ be an
eigenstate of a quantum system, ${\cal{A}}$ denote a geometric
configuration, in some cases ${\cal{A}}$ may be invariant to a
specific operation $\hat{O}$, we have then
\begin{equation}
 \hat{O}\Psi ({\cal{A}})=\Psi (\hat{O}{\cal{A}})=\Psi
 ({\cal{A}})\, .
\end{equation}
For example, when ${\cal{A}}$ is a regular octahedron (OCTA, see
Fig.~1) for a 6-body system, ${\cal{A}}$ is invariant to a
rotation about a 4-fold axis of the OCTA by $90^{\circ }$ together
with a cyclic permutation of the particles 1, 2, 3 and 4.
According to the representations of the operation on $\Psi $,
Eq.~(6) can always be written in a matrix form and appears as a
set of homogeneous linear equations. It is apparent that, whether
there exists nonzero solution of $\Psi({\cal{A}})$, in other
words, whether the state $\Psi$ is accessible to the configuration
${\cal{A}}$, depends on the inherent symmetric property of the
configuration. The symmetry imposes then a very strong constraint
on the eigenstate so that the $\Psi $ may be zero at ${\cal{A}}$.
It indicates that there may exist a specific kind nodal surface,
that is imposed by the intrinsic symmetry of the system (fixed at
body-frames) and independent of the dynamical property at all. One
usually refers such kind nodal surface (the second kind nodal
surface mentioned above) as inherent nodal surface (INS). It is
apparent whether the INS would appear is crucial to the low-lying
energy spectrum of a quantum system. The inherent nodal surface
structure analysis approach has then been proposed for few-body
electron and nucleon systems\cite{SPWbao1,SPWbao2,SPWbl}.

For a six-quark system, the orbital wave function may contain many
components. Each of them is specified by a set of quantum numbers
of inherent symmetry. Denoting $\lambda $ as a representation of
the permutation group \textit{S}$_{6}$, $i$\ a basis function of
this representation, $M$ the Z-component of $L$, and $g$ other
quantum numbers, we can express the components of the
wave-function as $F_{LMg}^{\lambda ,i}(1\cdot \cdot \cdot 6)$.
 Defining a body-frame, we have the
relation
\begin{equation}
F_{LMg}^{\lambda ,i}(1\cdot \cdot \cdot
6)=\sum_{Q}D_{QM}^{L}(-\gamma ,-\beta ,-\alpha )F_{LQg}^{\lambda
,i}(1^{\prime }\cdot \cdot \cdot 6^{\prime })\,,
\end{equation}
where $D_{QM}^{L}$ is an element of the matrix of rotation,
$\alpha ,\;\beta$ and $\gamma $ are the Euler angles specifying
the orientation of the body-frame, $Q$ denotes the projection of
$L$ along the third body-axis, \ $(1\cdot \cdot \cdot 6)$\ and
$(1^{\prime }\cdot \cdot \cdot 6^{\prime })$ stand for that the
coordinates are relative to the laboratory frame and to the
body-fixed frame, respectively. By analyzing the inherent nodal
surface structure of the system, we shall figure out which
components are advantageous to binding and thereby dominate the
low-lying states in what follows.

The symmetric operation on a system includes usually rotation,
space inversion, permutation of the particles, and so on.
According to the theory of space group, we can classify the
rotation axes into two kinds. If a geometric configuration
contains a $m$-fold axis of the first kind, the shape would be
invariant with respect to a rotation about the axis by the angle
$\frac{2\pi }{m}$. If it involves a $m$-fold axis of the second
kind, the shape would be invariant with respect to the rotation
together with a space inversion. In general, a configuration
containing at least one $m$-fold axis ($ m \geq $2) is called a
symmetric configuration. The spatial symmetry of a geometric
configuration is specified by the $m$-fold axes contained in the
configuration. For a six-quark system, there are many symmetric
configurations located everywhere in the coordinate space.
However, as a quantum system, not all the symmetric configurations
are allowed, since some of them might be prohibited by the
emergence of the INS.

As an example, let $OO^{\prime }$ be a 3-fold axis of a 6-body
configuration as shown in Fig.1, one can easily realize that a
rotation about $OO^{\prime }$ by $120^{\circ }$ is equivalent to
the cyclic permutation of particles 2, 5, 3 and that of 1, 4, 6,
respectively (see the figure). We have then
\begin{equation}
\hat{P}(253) \hat{P}(146)\hat{R}^{OO'}_{-120^{o}}F_{LQg}^{\lambda
,i}(1^{\prime }\cdots 6^{\prime })= F_{LQg}^{\lambda ,i}(1^{\prime
}\cdot \cdot \cdot 6^{\prime })\, .
\end{equation}
The crucial point is that $F_{LQg}^{\lambda ,i}$ is a basis
function of representations of both the spatial rotation group and
the permutation group. Hence, as the matrixes of representation of
the two operators are known, Eq.~(8) can be written in a matrix
form, and appears as a set of homogeneous linear equations. It is
well known that homogeneous equations might not have nonzero
solutions, which depends on whether the determinant of the matrix
is zero. The determinant depends on the $\lambda $ and $L$. Once
the determinant is nonzero, all the $F_{LQg}^{\lambda ,i}$ (all
$i$ and $Q$) must be zero at this symmetric configuration
disregarding the size of the shape and the permutation of the
particles at the vertexes of the shape. In other words, an
inherent nodal surface appears in these components of the
symmetric configuration. These components are then prohibited to
get access to this shape.

Incidentally, if the $m$-fold axis belongs to the second kind, the
constraint becomes

\begin{equation}
\hat{\mathbf{I}}\hat{P} \hat{R}F_{LQg}^{\lambda ,i}(1^{\prime
}\cdot \cdot \cdot 6^{\prime })= F_{LQg}^{\lambda ,i}(1^{\prime
}\cdot \cdot \cdot 6^{\prime })\,,
\end{equation}
where $\hat{\mathbf{I}}$  is the operator of space inversion.

Among all the symmetric configurations of a six-body system, the
one having the highest geometric symmetry (i.e., having the most
$m$-fold axes) is the regular octahedron (OCTA) as shown in Fig.1,
where three 4-fold axes, four 3-fold axes, and six 2-fold axes are
contained. Since each $m$-fold axis would cause a constraint,
evidently the OCTA is strongly constrained by symmetry. Only a
small portion of quantum wave-functions with specific $\lambda $
and $L$ can get access to it (i.e., the wave function is nonzero
at it). The accessibility of the OCTA\ turns out to be an
important point. Once the OCTA is accessible to a wave function,
all the symmetric configurations with a lower geometric symmetry
in the neighborhood of the OCTA (for example, the
prolonged-octahedron with the shape being prolonged along
$k^{\prime}$, the deformed-octahedron with the particles 1, 2, 3
and 4 illustrated in Fig.~1 forming a rectangle or a diamond) are
also accessible to the wave-function. Consequently, the
wave-function is able to distribute in a large domain including
the OCTA\ inside and free from the intervention of the INS. Such a
case is highly favorable to binding. Whereas if a wave function
contains an INS, its energy will increase drastically.

 \setlength{\unitlength}{1cm}
 \hspace{2cm}\begin{picture}(10,9)
            \put(1,3){\line(1,0){5}}
            \put(1,3){\line(1,2){1}}
            \multiput(2,5)(0.65,0){8}{\line(1,0){0.3}}

            \put(6,3){\line(1,2){1}}
            \put(2,5){\line(1,1){2}}
            \put(1,3){\line(3,4){3}}

            \put(6,3){\line(-1,2){2}}
            \put(7,5){\line(-3,2){3}}
            \put(1,3){\line(3,-2){3}}
            \put(6,3){\line(-1,-1){2}}
\put(1,3){\circle*{0.17}} \put(2,5){\circle*{0.17}}
\put(6,3){\circle*{0.17}} \put(7,5){\circle*{0.17}}
\put(4,7){\circle*{0.17}}\put(4,1){\circle*{0.17}}
\put(5.71,5.12){\circle{0.08}} \put(0.7,2.8){1} \put(6.1,2.8){2}
\put(7.1,4.9){3} \put(1.7,4.9){4} \put(4.1,7.1){5}
\put(4.1,0.7){6} \put(3.7,4){O} \put(5.3,5.1){O$^{\prime}$}
\put(4.1,7.8){k$^{\prime}$} \put(7.8,4.2){i$^{\prime}$}
            \multiput(4,1)(0.45,0.6){7}{\line(3,4){0.3}}

            \multiput(2,5)(0.35,-0.7){6}{\line(1,-2){0.2}}

            \put(4,4){\vector(0,1){4}}
            \put(4,4){\vector(1,0){4}}

\multiput(4,4)(0.54,0.36){3}{\line(3,2){0.36}}
            \put(5.5,5){\vector(3,2){2}}
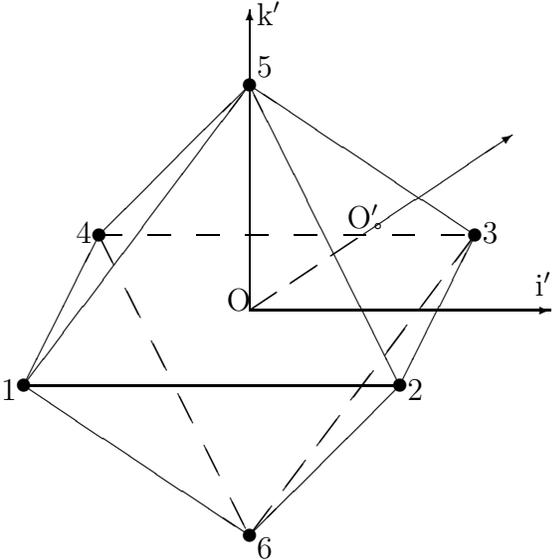
\put(1.2,-0.20){{Fig. 1} The regular octahedron.}
\end{picture}

\vspace*{5mm} \

The configuration with the second strongest geometric symmetry is
that the positions of the six quarks form a regular pentagon
pyramid(PENTA, shown as Fig.2). In an extreme case, the PENTA can
be C-PENTA, which corresponds to that with $h=0$ in Fig.~2. It is
evident that, if the C-PENTA is accessible, all the lower
symmetric configurations in its neighborhood (for instance, the
PENTA) are also accessible, since the restraints imposed by the
configuration are much weaker.

\vspace*{5mm}

\hspace{3cm}\begin{picture}(6,6)
            \put(1,3){\line(1,2){1}}
            \put(1,3){\line(1,-1){2}}
            \multiput(1,3)(0.6,0){8}{\line(1,0){0.3}}
            \put(2,5){\line(3,-1){3}}
            \put(5,4){\line(1,-2){1}}
            \put(6,2){\line(-3,-1){3}}

            \put(3,1){\line(1,6){0.5}}
            \put(1,3){\line(5,2){2.5}}
            \put(2,5){\line(3,-2){1.5}}
            \put(5,4){\line(-1,0){1.5}}
            \put(6,2){\line(-5,4){2.5}}
            \put(3.5,3){\line(0,1){1}}

            \put(5.5,3){\vector(1,0){1}}
            \put(3.5,4){\vector(0,1){1.5}}
 \put(0.7,2.8){5} \put(2.7,0.8){1} \put(6.2,1.8){2}
 \put(1.8,5.1){4} \put(3.6,4.1){6} \put(5.1,4.1){3}
 \put(3.5,2.7){O} \put(3.6,3.3){h} \put(3.6,5.5){k$^{\prime}$}
 \put(6.3,3.1){i$^{\prime}$}

 \put(1,3){\circle*{0.17}} \put(2,5){\circle*{0.17}}
\put(3.5,4){\circle*{0.17}} \put(5,4){\circle*{0.17}}
\put(6,2){\circle*{0.17}}\put(3,1){\circle*{0.17}}
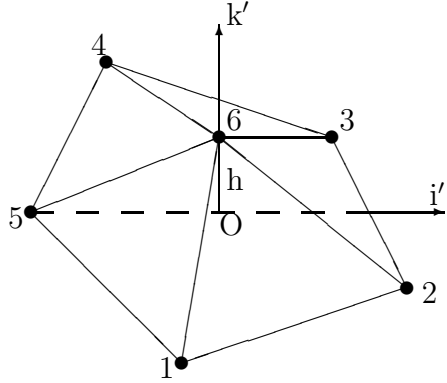
 \put(0,0){{Fig. 2} The regular pentagon pyramid.}
\end{picture}

\hspace{-5mm}\

It is evident that if a state does not contain a collective
excitation of rotation, i.e., the angular momentum L is zero, it
would be usually lower in energy than the state with $L\neq$0.
This is particularly true for the systems with a very small size
whose energy $E \propto \frac{L(L+1)}{r^2}$. On the experimental
side, it has been found that the ground states of 4 particles and
6 particles are always composed of components with L=0, while the
resonances with $L\geq$1 have quite higher excitation energies
(for a recent data compilation see Ref.\cite{Til02}). It is then
reasonable to assume that the low-lying states of the six-quark
system consist of components with $L=0$. Furthermore, the above
discussion shows that all the low-lying states in quantum
mechanics are mainly composed of nodeless components. Therefore,
if we limit the problem to the low-lying states, it is an
appropriate way to consider the S-wave nodeless components.
However, in view of the nucleon-nucleon collision, the P-wave
resonance may also be important.

Taking the way developed in Ref.\cite{SPWbl}, we obtain the
nodeless accessibility (i.e. number of the wave-functions) of the
configurations OCTA and C-PENTA with $L^{\pi}=0^{+}$, $1^{-}$ and
the possible spatial permutation symmetry S$_{6}$. The result is
that only the irreps
\begin{equation}
[f]_{O}=\{\lambda \} \in \{ [6 ], [4\,2], [2\, 2\, 2] \}
\end{equation}
of the S$_6$ for the S-wave states, and
\begin{equation}
\lbrack f\rbrack _{O} = \{ \lambda \}  \in \{ [ 5\,1] ,\; [
4\,1\,1], \; [ 3\,3] ,\; [ 3\,2\,1] ,\; [ 2\,2\,1\,1 ] \}
\end{equation}
for the P-wave resonance are nodeless accessible to both the OCTA
and the C-PENTA configurations.

It has been shown that the sources of INS  may exist in the
quantum states. Nonetheless, there are essentially
inherent-nodeless components of wave-functions (each with a
specific set of ($L^\pi, \lambda $)). The identification of these
particularly favorable components is a key to understand the
low-lying spectrum. Taking the standard method of irreducible
representation reduction, we get the quantum number spin($S$),
isospin($T$), total angular momentum($J$) and strangeness($s$) of
the S-wave states ($L^{\pi}=0^{+}$) with S$_{6}$ irreps $[6 ]$,
$[4\, 2]$, $[2\, 2 \, 2]$ and those of the P-wave states ($L^{\pi}
= 1 ^{-}$) with S$_{6}$ irreps $[5\, 1]$, $[4 \, 1 \, 1]$, $[3 \,
3]$, $[3 \, 2 \, 1]$, $[ 2\, 2\, 1\, 1 ]$. The obtained
accessibility is listed in Table 3 for the S-wave states and in
Table 4 for the P-wave states, respectively.

\section{The low-lying S-wave states}\label{sec:Study}

From the above discussion, we find that, when a wave-function
$\Psi_{LS\lambda}$ possesses quantum numbers
$(L^{\pi}\lambda)=(0^{+}[6])$, $(0^{+}[4\, 2])$, $(0^{+}[2\, 2\,
2])$, it can access both the OCTA\ and the C-PENTA. These and only
these $\Psi_{LS\lambda}$ are inherently nodeless components in the
two most important spatial configurations. They should then be the
dominant components of the S-wave low-lying states.

Because INS have no relation with dynamics, we can discuss the
low-lying states along the same line as that to discuss the
nucleus $^{6}$Li\cite{SPWbl}. Comparing the symmetry of the
dibaryons ($S_{6}$$\otimes$SU$_{CFS}$(18)) with that of the
nucleus $^{6}$Li ($S_{6}$$\otimes$SU$_{TS}(4)$), one can easily
infer that these two kind six-body systems have the same geometric
configurations. According to the energy spectrum of $^{6}$Li, we
can reach a conclusion that the states associating with orbital
symmetries $[4\, 2]$ and $[2\, 2\, 2]$ have lower energies than
those with only $[2\,2\,2 ]$ symmetry. Extending the results to
include the spatial symmetry $[6]$, we can compare the energies of
the states and deduce the low-lying ones.

\subsection{The states with strangeness 0}\label{sec:seq0}

From the above discussion we obtain that only the $(L^{\pi},
[f]_{O})= (0^{+},[6 ])$ and $(0^{+},[4\, 2])$ components are
nodeless accessible to the S-wave states. If the hidden-color
channel states are taken into account, the $(0^{+},[2\, 2\, 2 ])$
component is also allowed. From Table 3, one can realize that the
\{isospin, spin\} configurations $(T, S) = (0,1), \; (1, 0), \;
(1, 2),\; (2, 1)$ are the favorable ones since they are nodeless
accessible to all the four orbital-\{isospin, spin\}
configurations. Meanwhile, the states with $(T, S) = (0,3), \,
(3,0)$ may also be the low-lying dibaryon states since their
nodeless accessibility is the second largest one.

For the state with quantum number $(T,\; S)=(0, 1)$, both the
$[6]$ and $[4\, 2]$ components are available. Therefore, two
$T(J^{\pi})=0(1^{+})$ partner states would be generated. Each of
them is a specific mixture of the $[6 ]$ and $[4 \, 2]$
components. According to Harvey's result\cite{SPWharvey}, we know
one of them is the deuteron whose dominant orbital symmetry is
$[4\, 2 ]$. Of course, this is the lightest stable dibaryon.
Because of the interference between the $[6]$ and $[4 \, 2]$
components, there would be a gap between the partner states. The
interaction is spin-dependent and complicated, but we can expect
this gap to be large since even now there is no evidence for the
existence of the partner of deuteron. Furthermore, since the $[2\,
2\, 2 ]$ component is also accessible, there may exist a
hidden-color channel state with $(T, S) = (0, 1)$.

Referring to the order of states determined by considering the
color-magnetic interaction between quarks in quark
models\cite{SPWja}, we expect that, besides deuteron, the $(T,
S)=(1, 0)$ state with dominant orbital symmetry $[4\, 2]$ is the
second lightest one, since it has small spin and isospin. From
Table 2 we know that it is mainly a NN dibaryon. Since this state
obeys the Pauli principle of two-nucleon
systems($(-1)^{L+S+T}=-1$), it may decay into two protons if its
energy is not low enough. Then the possibility for it to be a
stable bound state may be not very large. Meanwhile, the
$T(J^{\pi})=1(0^{+})$ state also has a partner with dominant
orbital symmetry $[6]$. This state must have an energy much higher
than the $[4\, 2]$ component, it will be more easily to decay, so
we do not discuss it at all.

Our result for the states with low isospin and spin is similar to
that given in the quark-delocalization and color-screening model
(QDCSM)\cite{Wang92,Ping99}, where it was concluded that, besides
deuteron, $(T, S)=(1,\, 0)$ and $(1,\, 1)$ are the next two lowest
states and (1, 0) may be even a stable dibaryon if the interaction
is suitable. With regard to the state with $(T, S) = (0, 3)$, it
may correspond to the dibaryon $d^{\ast}$ proposed in the Los
Alamos potential model\cite{Gold89}, the QDCSM\cite{Wang95,
Ping99,Ping01,Ping02} and the Glauber multi-diffraction
model\cite{Wong98}. In addition, the presently proposed dibaryon
states with $(T, S) = (1, 2), \; (2,1)$ may correspond to the
narrow dibaryons proposed (with mass 2122~MeV, 2150~MeV,
$1956\pm6$~MeV, respectively) in the recent pp scattering
experiments\cite{SPWTa,SPWKh}. However the experimental data need
to be re-analyzed more carefully since similar experiment
accomplished most recently\cite{Brod02} does not provide any data
to support the result in the other mass region in
Ref.\cite{SPWTa}.

\subsection{States with strangeness $-5$ and $-6$}\label{sec:seq56}
It has been known that, when multi-quark system has a large
strangeness, the S-wave contribution is dominant. From Table 3, we
find that $(s, T, S)=(-5, \frac{1}{2}, 1)$, $(-5, \frac{1}{2}, 0)$
and $(-6, 0, 0)$ are the three lowest states. From Table 2, one
can obtain that the $(-6, 0, 0)$ component corresponds mainly to
the state $\Omega\Omega$. The lowest states with strangeness $-5$,
$(-5, \; \frac{1}{2}, 0)$, is $\Xi^{*}\Omega$ or a hidden-color
state. Another low-lying state, $(-5, \frac{1}{2}, 1)$, would be
$\Xi\Omega$ or the coupling of $\Xi\Omega$ and $\Xi^{*}\Omega$.

From Table 3, one can also recognize that $[\Omega\Omega]_{(0,
0^{+})}$ has a dominant orbital symmetry $[4\, 2]$. Its
hidden-color partner, which has the same quantum number $(T, S)$
but a different orbital symmetry $[6]$, might have such a high
energy that it cannot be stable. However, it can contribute to the
mass of $[\Omega\Omega]_{(0, 0^{+})}$ in experiment, if its mass
is not very large. $[\Xi^{*}\Omega]_{(1/2,0)}$ also has a
low-lying hidden-color partner and the result is the same as
$\Omega\Omega$.

After the possibility of the $s\! = \! -6$ dibaryon was proposed
in SU(3) chiral soliton model\cite{Kope908}, more sophisticated
calculations in the SU(3) chiral quark model\cite{ZY00} and
QDCSM\cite{Pang02} indicate that the $[\Omega \Omega
]_{(0,0^{+})}$ dibaryon may have binding energy $E_{b} = 45 \sim
115$~MeV. Recently, the color-singlet $\Xi^{*}\Omega\;(S=0, 3)$,
$\Xi\Omega$ and their coupling $\Xi^{*}\Omega$-$\Xi\Omega\;(S=1,
2)$ have also been studied in the chiral quark model\cite{LS00}.
It was found that, when $S=0$, $\Xi^{*}\Omega$ may be a dibaryon
with binding energy $E_{b}=80.0 \sim 92.4$~MeV. When $S=1$,
$\Xi\Omega$ may also be a dibaryon with binding energy $E_{b} =
26.2 \sim 32.9$~MeV. However, up to now, there is still no
evidence for the dibaryons with $S=2$ or $3$ to exist\cite{LS00}.
It indicates that $[\Omega\Omega]_{(0, 0)}$ and
$[\Xi^{*}\Omega]_{(1/2, 0)}$ are deeply bound dibaryon with narrow
widths, whereas $[\Xi\Omega]_{(1/2, 1)}$ and
$[\Xi\Omega$-$\Xi^{*}\Omega]_{(1/2, 1)}$ are weakly bound only if
the chiral field can provide an attraction among the baryons. The
present INS analysis result is consistent with these numerical
results very well.

It is worth to mention that some of these low-lying states may be
hidden-color states. Since the states have large strangeness, it
is hard to produce them in proton-proton collisions. However,
strangeness production can be enhanced in heavy ion
collisions\cite{Arms01}, especially at RHIC energies. The dibaryon
$[\Omega\Omega]_{(0, 0^{+})}$ has been suggested as one of the
possible signals of quark-gluon plasma(QGP) due to the large gluon
density and low energy threshold for $s\bar{s}$
formation\cite{Rafel82,Koch86,Pal01}. The dibaryons
$[\Omega\Omega]_{(0, 0^{+})}$ and $[\Xi^{*}\Omega]_{(1/2, 0^{+})}$
may then be observed in this kind of experiments after careful
discussion\cite{ZY00,Pang02}.

\subsection{States with strangeness $-4$}\label{sec:seq4}

Table 3 indicates that if $s=-4$, the states with $(T, S)=(0,
1),\; (1, 0),\; (1, 1)$ and $(1, 2)$ are the four low-lying
states. Considering the spin-dependent interaction again, we
expect the $(1, 0)$ state to be a stable one. According to Table
2, it is probably the $\Xi\Xi$ or $\Xi^{\ast} \Xi^{\ast}$ state.
Unlike the strangeless state $T(J^{\pi})=1(0^{+})$, $\Xi\Xi$ also
associates with orbital symmetries $\{5\, 1\}$ and $\{3\, 3\}$ in
P-wave resonance. This indicates that $\Xi\Xi$'s energy will not
be small enough. The recent SU(3) chiral quark model
calculation\cite{SPWswave} gave the similar result that it may be
a weakly bound state whose binding behavior depends on the
contribution of the chiral field. According to Table 2, the $(1,
0)$ state may be the $\Sigma^{*}\Omega$ state, too. However, until
now there is no numerical result of this state. We then suggest
that relevant calculation and discussion are necessary.

\subsection{States with strangeness $-1,$ $-2$ and $-3$}\label{sec:seq123}

From Table 3 one can easily recognize that almost all the states
with $s=-1, \, -2 , \, -3$ are associated with the orbital
symmetries $[6 ]$, $[4\, 2]$ and $[2\, 2\, 2]$ except for $(s, T,
S) = (-2, 0, 3)$. It is then difficult to rank them.

It has been well known that the six-quark system holds the
symmetry $S_{6}\otimes SU_{CFS}(18)$. According to group theory,
we can classify the quantum states basing on the group reduction
$S_{6}\otimes SU_{C}(3)\otimes SU_{S}(2)\otimes SU_{T}(2)\otimes
U_{s}(1)$, and obtain the quantum numbers of spin, isospin and
strangeness. The states determined in this way are always referred
to as in the ``symmetry" basis. On the other hand, we can classify
the states of the six-quark system in pairs of free baryons(e.g.
NN, $\Omega\Omega$ and so on) and hidden-color channel states with
explicit quantum numbers. These are referred to as in the
``physics" basis. The ``symmetry" basis can be transformed into
the ``physics" basis, and such a transformation causes couplings
among the states in ``symmetry" basis. When $s=0$, $-5$ and $-6$,
the couplings are simple, the states expressed in the ``physics"
basis present the main properties of those in the ``symmetry"
basis. Then the energies of the low-lying ``physics" states are
much different from each other. Nevertheless, when $s=-1$, $-2$
and $-3$, the couplings are so complicated that many low energy
states (with nodeless wave-functions) in the ``symmetry" basis
contribute to one ``physics" state. This makes it difficult to
separate each ``symmetry" state in experiment, since we can only
measure the ``physics" one. Meanwhile, due to the coupling, not
only low-lying states, but also higher energy states will combine
together. In this sense, the fact that the H-particle $(s=-2)$
predicted more than 20 years ago has not yet been observed in
experiments may be attributed to that several ``symmetry" states
with higher energies combine in the ``physics" state and make its
structure much more complicated. In fact, several theoretical and
experimental results in the same spirit have been proposed in the
last few years. For example, it has been found that when some
minor effects are included, the H dibaryon's mass will raise and
become above the $\Lambda\Lambda$
threshold\cite{SPWlattice,SPWQCM,SPWQCD}. However, right now we
can't conclude that there do not exist stable dibaryon when
$s=-1$, $-2$ and $-3$ until we get accurate quantitative results
about the couplings. After all, we should notice that this is an
important effect when we discuss the H particle and other dibaryon
with a medium strangeness.

\section{The low-lying P-wave states}\label{sec:Study2}

In view of the baryon-baryon collision, not only the bound states
but also the low-lying resonances are important. Moreover,
although we have found some states with low energies from the
above discussion, we do not have knowledge about whether they are
lower than the threshold of open channels. Hence, we can not
understand whether they are really bound. Considering the
resonance, one knows that the main feature is its width. If the
energy is higher than the threshold of open channels, the width is
too broad. Then the resonance can not be detected, and the
existence of such a resonance is meaningless. When the broadness
of the width is taken into account, the relative P-wave collision
is even more important than the S-wave collision for creating
narrow low-lying resonances, since there exists a centrifugal
barrier which would help to keep the orbital wave function in the
interior region. Therefore, in addition to the $L=0$ case having
been discussed above, we have to study the case of $L=1$ as
follows.

By calculating the determinants of the matrixes of the homogeneous
linear equations (as in Eqs.(8) and (9)) the inherently nodeless
components of a six-quark system with $L=1$ and $\pi _{dib}=-$\
can be identified. We obtain then the accessible orbital
symmetries as listed in Eq.(11). Accordingly, the favorable
components with $L^{\pi}=1^{-}$ and fulfilling all the
Eqs.(1)$\sim$(4), (8) and (9) are the three having symmetries
 $$\displaylines{\hspace*{2mm} \lbrack f\rbrack
_{O}\otimes \lbrack f\rbrack _{C}\otimes \lbrack f\rbrack _{FS}
\in \{ [ 5\,1] \otimes [ 2\,2\,2] \otimes [ 4\,2] ,\; [3\,3 ]
\otimes [2\,2\,2] \otimes [ 6], [3\,3] \otimes [2\,2\,2] \otimes
[4\,2] \} \,. \hfill{(12a)} \cr }$$
 If we go beyond the constraint in Eq.~(4),
we have the other possible building blocks of the P-wave resonance
$$\displaylines{\hspace*{2mm} \lbrack f\rbrack
_{O}\otimes \lbrack f\rbrack _{C}\otimes \lbrack f\rbrack _{FS}
\in \{ [4\,1\,1] \otimes [2\,2\,2 ] \otimes [4\,2] ,\; [3\,2\,1]
\otimes [2\,2\,2] \otimes [5\,1],\; \hfill{} \cr \hspace*{50mm}
[3\,2\,1] \otimes [2\,2\,2]\otimes [4\, 2], \; [2\,2\,1\,1]
\otimes [2\,2\,2]\otimes [4\,2] \} \,. \hfill{(12b)} \cr }$$
 This indicates that there may also exist some non-two-baryon states,
namely the hidden-color channel states.

Since $L=1$ states may always have energies larger than $L=0$
states, we can take the configuration (strangeness, isospin, spin)
= $(s, T, S)$ as the possible candidates of P-wave dibaryon only
if it accesses to the P-wave component but not to the S-wave one.
Looking over Table 4 and Table 3 together, one can recognize that
the configurations $(s, T, S)=(-6, 0, 1), \; (-4, 0, 0), \; (-2,
0,3), \;  (0, 0, 0), \; (0, 0, 2), \; (0, 2, 0), \; (0, 1, 3), \;
(0, 3, 1)$ and $(0, 3, 3)$ might be the low energy $L=1$ states.
Taking the Pauli principle into account, we find that the states
with the configuration $(s, T, S)=(-4, 0, 0)$, $(0, 0, 0)$, $(0,
0, 2)$, $(0, 2, 0)$, $(0, 1, 3)$, $(0, 3, 1)$ and $(0, 3, 3)$ can
decay into two baryons easily. One can thus obtain that there are
only two lowest P-wave states: $(T,S) = (0, 1)$ with $s=-6$ and
$(T, S) = (0, 3)$ with $s=-2$. Considering the characteristic of
the spin-orbital interaction among quarks (see for example
Ref.\cite{YZSD95}), one can identify that, when the $S$ and $L$
are fixed, the state with higher $J$ may have lower energy. The
possible $J^{\pi}$ of the above two states may then be $2^{-}$,
$4^{-}$, respectively. From Table 2, one can infer that they are
the $(\Omega \Omega)_{(0, 2^{-})}$, $(\Sigma ^{\ast}
\Sigma^{\ast})_{(0, 4^{-})}$ states and the hidden color channel
states with the same quantum numbers.

Even though the configuration $(s, T, S) = (0, 0, 1)$ is
accessible to the S-wave component (with multiplicity 5) and its
multiplicity to access the P-wave is only 1, it is still worth
being discussed, since bag-string model and chiral quark model
calculations\cite{Muld80,Kond87,SPWRGM} and some
experiments\cite{Bil938} show that there may exist dibaryon state
with exotic quantum numbers $L=1$, $T(J^{\pi})=0(0^{-})$ (denoted
as $d^{\prime}$). Since $d^{\prime}$ does not obey the Pauli
principle, it is a hidden-color channel state and does not simply
decay into two baryons. In the present INS analysis, among the
seven P-wave $[f]_O\otimes[f]_{FS}$ configurations, only $[3\, 2\,
1 ]\otimes[5\, 1]$ is the nodeless component accessible to the
$(s, T, S)=(0, 0, 1)$ state. The $d^{\prime}$ may then not have a
low energy due to the emergence of INS. Once $d^{\prime}$ has low
energy, it must hold a dominant component with orbital symmetry
$[f]_O=[3\, 2\, 1 ]$. However, this orbital symmetry has not yet
been discussed in the chiral quark model\cite{SPWRGM}. Moreover,
the QDCSM calculation indicates that the dibaryon $d^{\prime}$ may
involve more complicated structure\cite{Ping00}. Then, more
careful investigations may be interesting to check the existence
of $d^{\prime}$ further.

Since spin-dependent quark-quark interactions can lower the energy
of the excited s$^{4}$p$^{2}$ state and raise the energy of the
s$^{6}$ state\cite{SPWRGM}, the dibaryon states $d^{*}$ (with
$T(J^{\pi})=0(3^{+})$)\cite{Wang95,Ping99,Gold89,Ping01,Ping02,Wong98}
and $d^{\prime}$ are considered to have energies lower than S-wave
$(T, S)=(1, 0)$ state. In view of that all the $s=0$ states are
very near the $\pi\pi$ threshold, one could easily reach a
conclusion that, except for the deuteron, all of the S-wave
dibaryons with $s=0$ have energies higher than $d^{\prime}$.
However, the inelastic scattering experiments $pp\Rightarrow p
\pi^{+}X$\cite{SPWTa}, $pp\Rightarrow X\gamma\gamma$\cite{SPWKh}
and quasielastic-charge-exchange experiment $p \alpha \Rightarrow
n ppd,\, pppnn$\cite{SPWBl} give very slight evidence for narrow
dibaryon $d^{\prime}$, and the quantum numbers $(T, J^{\pi}) = (0,
0^{-})$ for $d^{\prime}$ should be carefully
reexamined\cite{SPWTa}. Furthermore, the most recent $p p
\Rightarrow pp\pi^{+} \pi^{-}$ measurements\cite{Brod02} indicates
that the so called dibaryon $d^{\prime}$ may not exist at all. The
result of very small accessibility obtained in the present
analysis may provide a reason to understand such a contradiction.
In contrast, the large accessibility of the $(T,S)=(0,3)$ state
discussed in the last section provides a clue that the $d^{\ast}$
dibaryon may be observed in experiment.

Referring the states with large accessibility as the states
favorable to binding, we obtain finally the candidates of
low-lying dibaryons from the above discussion. The results are
listed in Table 5.

\section{Summary and Remarks}\label{sec:Summ}

We have studied a kind exotic QCD state, namely the dibaryon
states, in this paper. Taking the dibaryons as six-quark clusters
with $u$, $d$ and $s$ quarks, we have discussed the symmetries of
the states. After investigating the inherent nodal surface of the
system based on symmetry analysis, we obtain the quantum numbers
of the low-lying S-wave and P-wave states. Meanwhile the
hidden-color channel states are discussed. It shows that although
several states with strangeness $s=0$ may have low energy, there
is probably no stable S-wave dibaryon except deuteron. If the
strangeness is large, however, there may exist stable dibaryon
states. $[\Omega\Omega]_{(0, 0^{+})}$, $[\Xi^{*}\Omega]_{(1/2,
0^{+})}$, $[\Xi^{*}\Omega]_{(1/2, 1^{+})}$ and $[\Xi
\Omega]_{(1/2, 1^{+})}$ are proposed to be the candidates of
stable dibaryons. $[\Xi\Omega$-$\Xi^{*}\Omega]_{(1/2, 1^{+})}$ and
$[\Xi\Xi]_{(1, 0^{+})}$ are low-lying states, too. However, they
may be not stable enough. We also find that two important P-wave
resonances $(s, T, S) = (-6, 0, 1)$, $(-2, 0, 3)$ may be stable
and have low energy. Nevertheless, the $d^{\prime}$ with $(s, T,
S) = (0, 0, 1)$ is not a favorite candidate of dibaryon in the
present analysis. It is evident that these results agree very well
with those obtained in various numerical calculations and
experimental observations. In the case that $s=-1$, $-2$ and $-3$,
it is difficult to determine the low-lying states since many such
kind states are associated with each other. This kind mixing may
be the reason for that the H-particle has not yet been observed in
experiment convincingly.

Combining our present analysis and the previous numerical
calculations, we conclude that the S-wave dibaryon states
$[\Omega\Omega]_{(0,0)}$ and $[\Xi^{*}\Omega]_{(1/2,0)}$ may be
observed in the RHIC experiments since they are the stable
low-lying states. Meanwhile, the P-wave states
$[\Omega\Omega]_{(0,2^{-})}$ and
$[\Sigma^{*}\Sigma^{*}]_{(0,4^{-})}$ may also be observed as the
dibaryon candidates. Since all of them may contain hidden-color
channel state couplings, we propose that observing the states with
hidden-color channels may be an appropriate branch to explore
dibaryon states. In addition, the dibaryon $d^{\ast}$ may also be
observed in experiment.

Finally, it is remarkable that the presently proposed INS analysis
approach is just a mathematical realization of the intrinsic
property of the underlying physics. No dynamics is included at
all. The results obtained are then model independent. However, it
can give only qualitative predictions. To obtain concrete
knowledge about dibaryon states, for instance their binding
energies, decay widths, one should implement dynamical
calculations definitely. Therefore, combining the dynamical
calculation and INS analysis together is much more efficient in
exploring dibaryons and other multi-quark cluster states.

\bigskip

% Acknowledgements
This work is supported by the National Science Foundation of China
with Grant Nos. 10075002, 10135030, 19875001, 90103028. One of the
authors (YXL) thanks also the support by the Major State Key Basis
Research Programme under Contract No. G2000077400 and the
Foundation for University Key Teacher by the Ministry of
Education, China.

\begin{table}[htbp]\begin{center}
\caption{\small{The flavor-spin symmetries corresponding to each
possible orbital symmetry with color singlet restriction.}}

\vspace*{2mm}
\begin{tabular} {|c|l|} \hline
Orbital (S$_{6})$ &\ \ \ \ \ \ \ \ \ \ SU(6)$_{FS}$ \\
\hline $[6]$       &$[3\, 3]^{*}$ \\
\hline $[5\,1]$    &$[4\,2]^{*}$, $[3\, 2\, 1]$ \\
\hline $[4\,2]$    &$[5\,1]^{*}$, $[4\,1\,1]$, $[3\,3]^{*}$,
                    $[3\,2\,1]$, $[2\,2\,1\,1]$ \\
\hline $[4\,1\,1]$  &$[4\,2]^{*}$, $[4\,1\,1]$, $[3\,2\,1]$, $[2\,2\,1\,1]$ \\
\hline $[3\,3]$    &$[6]^{*}$, $[4\,2]^{*}$, $[3\,1^{3}]$, $[2\,2\,2]$ \\
\hline $[3\,2\,1[$  &$[5\,1]^{*}$, $[4\,2]^{*}$, $[4\,1\,1]$,
                   $[3\,1^{3}]$, $[3\,2\,1]^{2}$, $[2\,2\,1\,1]$, $[2\,1^{4}]$ \\
\hline $[3\,1^{3}[$ &$[4\,1\,1]$, $[3\,3]^{*}$, $[3\,2\,1]$,
                      $[3\,1^{3}]$, $[2\,2\,1\,1]$ \\
\hline $[2\,2\,2]$   &$[4\,1\,1]$, $[3\,3]^{*}$, $[3\,1^{3}]$,
                      $[2\,2\,1\,1]$, $[1^{6}]$ \\
\hline $[2\,2\,1\,1]$  &$[4\,2]^{*}$, $[3\, 1^{3}]$, $[2\,2\,2]$ \\
\hline $[2\,1^{4}]$ &$[3\,2\,1]$, $[2\,2\,1\,1]$ \\
\hline $[1^{6}]$  &$[2\,2\,2]$ \\
\hline
\end{tabular}
\end{center}
\end{table}

\renewcommand{\baselinestretch}{1.5}

\begin{table}[htbp]
\begin{center}
\caption{\small{Classification of the states with asterisks in
table 1 }}

\vspace{2mm }

\begin{tabular}
{|c|c|p{310pt}|} \hline strange& states&
(isospin, spin) \\
\hline \raisebox{-12.0ex}[0cm][0cm]
        {0}&     N$^{2}$& (0,\ 0), (0,\ 1), (1,\ 0), (1,\ 1) \\
\cline{2-3}&$\Delta^{2}$& (0,\ 0), (0,\ 1), (0,\ 2), (0,\ 3), (1,\
0), (1,\ 1), (1,\ 2), (1,\ 3), \par
  (2,\ 0), (2,\ 1), (2,\ 2), (2,\ 3), (3,\ 0), (3,\ 1), (3,\ 2), (3,\ 3), \\
\cline{2-3}&  N$\Delta $& (1,\ 1)$^{2}$, (1,\ 2)$^{2}$, (2,\ 1)$^{2}$, (2,\ 2)$^{2}$ \\
\cline{2-3}&          CC& (0,\ 0), (0,\ 1), (0,\ 2), (0,\ 3), (1,\
0), (1,\ 1)$^{2}$, (1,\ 2)$^{2}$, \par
 (1,\ 3), (2,\ 0), (2,\ 1)$^{2}$, (2,\ 2), (3,\ 0), (3,\ 1), \\
\hline \raisebox{-20.00ex}[0cm][0cm]
       {$-1$}& N$\Lambda $& (1/2,\ 0)$^{2}$, (1/2,\ 1)$^{2}$ \\
\cline{2-3}& N$\Sigma  $& (3/2,\ 0)$^{2}$, (3/2,\ 1)$^{2}$, (1/2,\ 0)$^{2}$, (1/2,\ 1)$^{2}$ \\
\cline{2-3}& N$\Sigma^{*}  $& (3/2,\ 1)$^{2}$, (3/2,\ 2)$^{2}$, (1/2,\ 1)$^{2}$, (1/2,\ 2)$^{2}$ \\
\cline{2-3}& $\Delta \Lambda $&(3/2,\ 1)$^{2}$, (3/2,\ 2)$^{2}$ \\
\cline{2-3}& $\Delta \Sigma  $&(5/2,\ 1)$^{2}$, (5/2,\ 2)$^{2}$,
(3/2,\ 1)$^{2}$,
(3/2,\ 2)$^{2}$, (1/2,\ 1)$^{2}$, (1/2,\ 2)$^{2}$   \\
\cline{2-3}& $\Delta \Sigma^{*}  $&(5/2,\ 0)$^{2}$,(5/2,\
1)$^{2}$, (5/2,\ 2)$^{2}$,(5/2,\ 3)$^{2}$, (3/2,\ 0)$^{2}$, (3/2,\
1)$^{2}$,\par (3/2,\ 2)$^{2}$, (3/2,\ 3)$^{2}$, (1/2,\ 0)$^{2}$,
(1/2,\ 1)$^{2}$, (1/2,\ 2)$^{2}$, (1/2,\ 3)$^{2}$ \\
\cline{2-3}&          CC& (1/2,\ 0)$^{3}$, (1/2,\ 1)$^{6}$, (1/2,\
2)$^{5}$, (1/2,\ 3)$^{2}$, (3/2,\ 0)$^{3}$, (3/2,\ 1)$^{6}$, \par
(3/2,\ 2)$^{4}$, (3/2,\ 3), (5/2,\ 0)$^{2}$,
(5/2,\ 1)$^{3}$, (5/2,\ 2), (5/2,\ 3)\\
\hline \raisebox{-21.00ex}[0cm][0cm]
       {$-2$}&$\Lambda^{2}$&(0,\ 0), (0,\ 1) \\
\cline{2-3}&$\Sigma^{2} $&(0,\ 0), (0,\ 1), (1,\ 0), (1,\ 1) \\
\cline{2-3}&$\Sigma^{*2} $&(0,\ 0), (0,\ 1), (0,\ 2), (0,\ 3), (1,\ 0), (1,\ 1), (1,\ 2), (1,\ 3) \\
\cline{2-3}&$\Sigma^{*}\Sigma $&(0,\ 1)$^{2}$, (0,\ 2)$^{2}$, (1,\ 1)$^{2}$, (1,\ 2)$^{2}$ \\
\cline{2-3}&      N$\Xi $&(0,\ 0)$^{2}$, (0,\ 1)$^{2}$, (1,\ 0)$^{2}$, (1,\ 1)$^{2}$ \\
\cline{2-3}&      N$\Xi^{*}$&(0,\ 1)$^{2}$, (0,\ 2)$^{2}$, (1,\ 1)$^{2}$, (1,\ 2)$^{2}$ \\
\cline{2-3}& $\Delta \Xi$&(1,\ 1)$^{2}$, (1,\ 2)$^{2}$, (2,\ 1)$^{2}$, (2,\ 2)$^{2}$  \\
\cline{2-3}& $\Delta \Xi^{*}$&(1,\ 0)$^{2}$, (1,\ 1)$^{2}$, (1,\
2)$^{2}$, (1,\ 3)$^{2}$, (2,\ 0)$^{2}$, (2,\ 1)$^{2}$, (2,\
2)$^{2}$, (2,\ 3)$^{2}$  \\
 \cline{2-3}& CC &(0,\ 0), (0,\
1)$^{4}$, (0,\ 2)$^{4}$, (0,\ 3), (1,\ 0)$^{7}$, (1,\ 1)$^{4}$,
\par (1,\ 2)$^{9}$, (1,\ 3)$^{2}$,  (2,\ 0)$^{5}$, (2,\ 1)$^{9}$, (2,\ 2)$^{5}$, (2,\ 3)$^{2}$ \\
\hline
\end{tabular}
\end{center}
\end{table}

\begin{table}
{\hspace*{1cm} \small (Table 2 continued)}
\begin{center}
\begin{tabular}{|c|c|p{290pt}|}
\hline strange& states&
(isospin, spin) \\
 \hline \raisebox{-22.00ex}[0cm][0cm]
       {$-3$}&  N$\Omega $&(1/2,\ 1)$^{2}$, (1/2,\ 2)$^{2}$ \\
\cline{2-3}& $\Delta \Omega $&(3/2,\ 0)$^{2}$, (3/2,\ 1)$^{2}$, (3/2,\ 2)$^{2}$, (3/2,\ 3)$^{2}$ \\
\cline{2-3}& $\Lambda \Xi$&(1/2,\ 0)$^{2}$, (1/2,\ 1)$^{2}$ \\
\cline{2-3}& $\Lambda \Xi^{*}$&(1/2,\ 1)$^{2}$, (1/2,\ 2)$^{2}$ \\
\cline{2-3}& $\Sigma \Xi $&(1/2,\ 0)$^{2}$, (1/2,\ 1)$^{2}$, (3/2,\ 0)$^{2}$, (3/2,\ 1)$^{2}$   \\
\cline{2-3}& $\Sigma^{*} \Xi $&(1/2,\ 1)$^{2}$, (1/2,\ 2)$^{2}$, (3/2,\ 1)$^{2}$, (3/2,\ 2)$^{2}$   \\
\cline{2-3}& $\Sigma \Xi^{*} $&(1/2,\ 1)$^{2}$, (1/2,\ 2)$^{2}$, (3/2,\ 1)$^{2}$, (3/2,\ 2)$^{2}$   \\
\cline{2-3}& $\Sigma^{*} \Xi^{*} $&(1/2,\ 0)$^{2}$, (1/2,\
1)$^{2}$, (1/2,\ 2)$^{2}$, (1/2,\ 3)$^{2}$,
\par (3/2,\ 0)$^{2}$, (3/2,\ 1)$^{2}$, (3/2,\ 2)$^{2}$, (3/2,\ 3)$^{2}$
\\
\cline{2-3}&            CC&(1/2,\ 0)$^{6}$, (1/2,\ 1)$^{9}$,
(1/2,\ 2)$^{5}$, (1/2,\ 3),
\par (3/2,\ 0)$^{5}$, (3/2,\ 1)$^{6}$, (3/2,\ 2)$^{3}$, (3/2,\ 3)$^{2}$ \\
 \hline
\raisebox{-13.00ex}[0cm][0cm]
       {$-4$}& $\Lambda \Omega $&(0,\ 1)$^{2}$, (0,\ 2)$^{2}$ \\
\cline{2-3}&  $\Sigma \Omega $&(1,\ 1)$^{2}$, (1,\ 2)$^{2}$ \\
\cline{2-3}&  $\Sigma^{*} \Omega $&(1,\ 0)$^{2}$, (1,\ 1)$^{2}$, (1,\ 2)$^{2}$, (1,\ 3)$^{2}$\\
\cline{2-3}& $\Xi ^{2}       $&(0,\ 0), (0,\ 1), (1,\ 0), (1,\ 1)  \\
\cline{2-3}& $\Xi^{*}\Xi       $&(0,\ 1)$^{2}$, (0,\ 2)$^{2}$, (1,\ 1)$^{2}$, (1,\ 2)$^{2}$  \\
\cline{2-3}& $\Xi ^{*2}       $&(0,\ 0), (0,\ 1), (0,\ 2), (0,\ 3), (1,\ 0), (1,\ 1), (1,\ 2), (1,\ 3)\\
\cline{2-3}&                CC&(0,\ 0)$^{2}$, (0,\ 1)$^{4}$, (0,\
2)$^{4}$, (1,\ 0)$^{3}$,
(1,\ 1)$^{6}$, (1,\ 2)$^{3}$, (1,\ 3)\\
\hline \raisebox{-4ex}[0cm][0cm]
       {$-5$}& $\Omega \Xi$&(1/2,\ 1)$^{2}$, (1/2,\ 2)$^{2}$   \\
\cline{2-3}& $\Omega \Xi^{*}$&(1/2,\ 0)$^{2}$, (1/2,\ 1)$^{2}$, (1/2,\ 2)$^{2}$, (1/2,\ 3)$^{3}$   \\
\cline{2-3}&           CC&(1/2,\ 0)$^{2}$, (1/2,\ 1)$^{3}$, (1/2,\ 2)\\
\hline \raisebox{-2ex}[0cm][0cm]
       {$-6$}& $\Omega^{2}$&(0,\ 0), (0,\ 1), (0,\ 2), (0,\ 3)  \\
 \cline{2-3}&           CC&(0,\ 0), (0,\ 1) \\
\hline
\end{tabular}
\end{center}
\end{table}

\begin{table}[htbp]
\begin{center}
\caption{\small {Multiplicity of the accessibility of the S-wave
nodeless components $[f]_O \otimes [f]_{FS}$ }}

\begin{footnotesize}
\begin{tabular}
{|c|c|c|c|c|c|c|} \hline
%\multicolumn{3}{|c|}{\textbf{states}} &
%\multicolumn{7}{|c|}{\textbf{P-wave nodeless components}} &
%\multicolumn{4}{|c|}{\textbf{S-wave nodeless components}}  \\
\hline $s$& $S$& $T$& $[6] \otimes [3\, 3]$ & $[4\, 2] \otimes
[5\, 1]$ & $[4\, 2] \otimes [3\, 3]$ & $[2\, 2\, 2] \otimes [3\,
3]$ \\  \hline
 \raisebox{-32.50ex}[0cm][0cm]{0}& 0& 0&  0& 0& 0& 0 \\
\cline{2-7}
 & 0& 1& 1& 1& 1& 1 \\
\cline{2-7}
 & 0& 2& 0& 0& 0& 0 \\
\cline{2-7}
 & 0 & 3 & 1 & 0 & 1 & 1 \\
\cline{2-7}
 & 1 & 0 & 1 & 1 & 1 & 1 \\
\cline{2-7}
 & 1 & 1 & 0 & 1 & 0 & 0 \\
\cline{2-7}
 & 1 & 2 & 1 & 1 & 1 & 1 \\
\cline{2-7}
 & 1 & 3 & 0 & 0 & 0 & 0 \\
\cline{2-7}
 & 2 & 0 & 0 & 0 & 0 & 0 \\
\cline{2-7}
 & 2 & 1 & 1 & 1 & 1 & 1 \\
\cline{2-7}
 & 2 & 2 & 0 & 1 & 0 & 0 \\
\cline{2-7}
 & 2 & 3 & 0 & 1 & 0 & 0 \\
\cline{2-7}
 & 3 & 0 & 1 & 0 & 1 & 1 \\
\cline{2-7}
 & 3 & 1 & 0 & 0 & 0 & 0 \\
\cline{2-7}
 & 3 & 2 & 0 & 1 & 0 & 0 \\
\cline{2-7}
 & 3 & 3 & 0 & 0 & 0 & 0 \\
\hline \raisebox{-24.50ex}[0cm][0cm]{$-1$}& 0& 1/2& 1& 2& 1& 1 \\
\cline{2-7}
 & 0 & 3/2 & 1 & 1 & 1 & 1 \\
\cline{2-7}
 & 0 & 5/2 & 1 & 0 & 1 & 1 \\
\cline{2-7}
 & 1 & 1/2 & 2 & 3 & 2 & 2 \\
\cline{2-7}
 & 1 & 3/2 & 2 & 3 & 2 & 2 \\
\cline{2-7}
 & 1 & 5/2 & 1 & 1 & 1 & 1 \\
\cline{2-7}
 & 2 & 1/2 & 2 & 1 & 2 & 2 \\
\cline{2-7}
 & 2 & 3/2 & 1 & 3 & 1 & 1 \\
\cline{2-7}
 & 2 & 5/2 & 0 & 2 & 0 & 0 \\
\cline{2-7}
 & 3 & 1/2 & 1 & 0 & 1 & 1 \\
\cline{2-7}
 & 3 & 3/2 & 0 & 1 & 0 & 0 \\
\cline{2-7}
 & 3 & 5/2 & 0 & 1 & 0 & 0 \\
\hline
\end{tabular}
\end{footnotesize}
\end{center}
%\label{tab1}
\end{table}

\begin{table}[htbp]
{\hspace*{3cm} \small (Table 3 continued)}
\begin{center}
\begin{footnotesize}
\begin{tabular}
{|c|c|c|c|c|c|c|} \hline ${s}$ & ${S}$ & ${T}$ & $[6 ] \otimes
[3\, 3]$ & $[4\, 2] \otimes [5\, 1]$ &
$[4\, 2] \otimes [3\, 3]$ & $[2\, 2\, 2] \otimes [3\, 3]$ \\
\hline
\raisebox{-24.50ex}[0cm][0cm]{$-2$}& 0& 0& 2& 2& 2& 2 \\
\cline{2-7}
 & 0 & 1 & 1 & 2 & 1 & 1 \\
\cline{2-7}
 & 0 & 2 & 2 & 1 & 2 & 2 \\
\cline{2-7}
 & 1 & 0 & 1 & 2 & 1 & 1 \\
\cline{2-7}
 & 1 & 1 & 4 & 5 & 4 & 4 \\
\cline{2-7}
 & 1 & 2 & 1 & 2 & 1 & 1 \\
\cline{2-7}
 & 2 & 0 & 2 & 1 & 2 & 2 \\
\cline{2-7}
 & 2 & 1 & 2 & 3 & 2 & 2 \\
\cline{2-7}
 & 2 & 2 & 1 & 2 & 1 & 1 \\
\cline{2-7}
 & 3 & 0 & 0 & 0 & 0 & 0 \\
\cline{2-7}
 & 3 & 1 & 1 & 1 & 1 & 1 \\
\cline{2-7}
 & 3 & 2 & 0 & 1 & 0 & 0 \\
\hline \raisebox{-15.50ex}[0cm][0cm]{$-3$}& 0& 1/2& 1& 2& 1& 1 \\
\cline{2-7}
 & 0 & 3/2 & 2 & 1 & 2 & 2 \\
\cline{2-7}
 & 1 & 1/2 & 3 & 4 & 3 & 3 \\
\cline{2-7}
 & 1 & 3/2 & 2 & 3 & 2 & 2 \\
\cline{2-7}
 & 2 & 1/2 & 2 & 3 & 2 & 2 \\
\cline{2-7}
 & 2 & 3/2 & 1 & 3 & 1 & 1 \\
\cline{2-7}
 & 3 & 1/2 & 0 & 1 & 0 & 0 \\
\cline{2-7}
 & 3 & 3/2 & 1 & 1 & 1 & 1 \\
\hline \raisebox{-15.50ex}[0cm][0cm]{$-4$}& 0& 0& 0& 0& 0& 0 \\
\cline{2-7}
 & 0 & 1 & 2 & 1 & 2 & 2 \\
\cline{2-7}
 & 1 & 0 & 2 & 2 & 2 & 2 \\
\cline{2-7}
 & 1 & 1 & 1 & 2 & 1 & 1 \\
\cline{2-7}
 & 2 & 0 & 0 & 2 & 0 & 0 \\
\cline{2-7}
 & 2 & 1 & 1 & 3 & 1 & 1 \\
\cline{2-7}
 & 3 & 0 & 0 & 1 & 0 & 0 \\
\cline{2-7}
 & 3 & 1 & 0 & 1 & 0 & 0 \\
\hline
\end{tabular}
\end{footnotesize}
\end{center}
%\label{tab1}
\end{table}

\begin{table}[htbp]
{\hspace*{3cm} \small (Table 3 continued)}
\begin{center}
\begin{footnotesize}
\begin{tabular}
{|c|c|c|c|c|c|c|} \hline ${s}$ & ${S}$ & ${T}$ & $[6 ] \otimes
[3\, 3]$ & $[4\, 2] \otimes [5\, 1]$ &
$[4\, 2] \otimes [3\, 3]$ & $[2\, 2\, 2] \otimes [3\, 3]$ \\
\hline
\raisebox{-6.50ex}[0cm][0cm]{$-5$}& 0& 1/2& 1& 0& 1& 1 \\
\cline{2-7}
 & 1 & 1/2 & 1 & 1 & 1 & 1 \\
\cline{2-7}
 & 2 & 1/2 & 0 & 2 & 0 & 0 \\
\cline{2-7}
 & 3 & 1/2 & 0 & 1 & 0 & 0 \\
\hline \raisebox{-6.50ex}[0cm][0cm]{$-6$}& 0& 0& 1& 0& 1& 1 \\
\cline{2-7}
 & 1 & 0 & 0 & 0 & 0 & 0 \\
\cline{2-7}
 & 2 & 0 & 0 & 1 & 0 & 0 \\
\cline{2-7}
 & 3 & 0 & 0 & 0 & 0 & 0 \\
\hline
\end{tabular}
\end{footnotesize}
\end{center}
%\label{tab1}
\end{table}

\begin{table}[htbp]
\begin{center}
\caption{\small {Multiplicity of the accessibility of the P-wave
nodeless components $[f]_O \otimes [f]_{FS}$ }}

\begin{footnotesize}
\begin{tabular}
{|c|c|c|c|c|c|c|c|c|c|} \hline
%\multicolumn{3}{|c|}{\textbf{states}} &
%\multicolumn{7}{|c|}{\textbf{P-wave nodeless components}} &
%\multicolumn{4}{|c|}{\textbf{S-wave nodeless components}}  \\
\hline ${s}$ & ${S}$ & ${T}$ & $\begin{array}{c} [5\, 1] \otimes
\\ \noalign{\vskip -2mm} [4\, 2] \\ \end{array} $ & $\begin{array}{c}
[4\,1\, 1] \\ \noalign{\vskip -2mm} \otimes [4
\, 2] \\ \end{array} $ & $\begin{array}{c} [3\, 3] \\
\noalign{\vskip -2mm} \otimes [6] \\ \end{array} $ &
$\begin{array}{c} [3\, 3] \\ \noalign{\vskip -2mm} \otimes [4 \,
2] \\ \end{array} $ & $\begin{array}{c} [3\, 2\, 1] \\
\noalign{\vskip -2mm} \otimes [5\, 1] \\ \end{array} $ &
$\begin{array}{c} [3\,2\, 1] \\ \noalign{\vskip -2mm} \otimes [4
\, 2] \\ \end{array} $ &
$\begin{array}{c} [2\,2\,1\,1] \\ \noalign{\vskip -2mm} \otimes [4\, 2] \\
\end{array} $ \\
\hline \raisebox{-33.50ex}[0cm][0cm]{0}& 0& 0& 1& 1& 1& 1& 0& 1& 1 \\
\cline{2-10}
 & 0 & 1 & 0 & 0 & 0 & 0 & 1 & 0 & 0 \\
\cline{2-10}
 & 0 & 2 & 1 & 1 & 0 & 1 & 0 & 1 & 1 \\
\cline{2-10}
 & 0 & 3 & 0 & 0 & 0 & 0 & 0 & 0 & 0 \\
\cline{2-10}
 & 1 & 0 & 0 & 0 & 0 & 0 & 1 & 0 & 0 \\
\cline{2-10}
 & 1 & 1 & 2 & 2 & 1 & 2 & 1 & 2 & 2 \\
\cline{2-10}
 & 1 & 2 & 1 & 1 & 0 & 1 & 1 & 1 & 1 \\
\cline{2-10}
 & 1 & 3 & 1 & 1 & 0 & 1 & 0 & 1 & 1 \\
\cline{2-10}
 & 2 & 0 & 1 & 1 & 0 & 1 & 0 & 1 & 1 \\
\cline{2-10}
 & 2 & 1 & 1 & 1 & 0 & 1 & 1 & 1 & 1 \\
\cline{2-10}
 & 2 & 2 & 1 & 1 & 1 & 1 & 1 & 1 & 1 \\
\cline{2-10}
 & 2 & 3 & 0 & 0 & 0 & 0 & 1 & 0 & 0 \\
\cline{2-10}
 & 3 & 0 & 0 & 0 & 0 & 0 & 0 & 0 & 0 \\
\cline{2-10}
 & 3 & 1 & 1 & 1 & 0 & 1 & 0 & 1 & 1 \\
\cline{2-10}
 & 3 & 2 & 0 & 0 & 0 & 0 & 1 & 0 & 0 \\
\cline{2-10}
 & 3 & 3 & 0 & 0 & 1 & 0 & 0 & 0 & 0 \\
\hline \raisebox{-24.50ex}[0cm][0cm]{$-1$}& 0& 1/2& 2& 2& 1&
2& 2& 2& 2 \\
\cline{2-10}
 & 0 & 3/2 & 2 & 2 & 0 & 2 & 1 & 2 & 2 \\
\cline{2-10}
 & 0 & 5/2 & 1 & 1 & 0 & 1 & 0 & 1 & 1 \\
\cline{2-10}
 & 1 & 1/2 & 4 & 4 & 1 & 4 & 3 & 4 & 4 \\
\cline{2-10}
 & 1 & 3/2 & 4 & 4 & 1 & 4 & 3 & 4 & 4 \\
\cline{2-10}
 & 1 & 5/2 & 2 & 2 & 0 & 2 & 1 & 2 & 2 \\
\cline{2-10}
 & 2 & 1/2 & 3 & 3 & 0 & 3 & 1 & 3 & 3 \\
\cline{2-10}
 & 2 & 3/2 & 3 & 3 & 1 & 3 & 3 & 3 & 3 \\
\cline{2-10}
 & 2 & 5/2 & 1 & 1 & 1 & 1 & 2 & 1 & 1 \\
\cline{2-10}
 & 3 & 1/2 & 1 & 1 & 0 & 1 & 0 & 1 & 1 \\
\cline{2-10}
 & 3 & 3/2 & 1 & 1 & 0 & 1 & 1 & 1 & 1 \\
\cline{2-10}
 & 3 & 5/2 & 0 & 0 & 1 & 0 & 1 & 0 & 0 \\
\hline
\end{tabular}
\end{footnotesize}
\end{center}
%\label{tab1}
\end{table}

\begin{table}[htbp]
{\hspace*{1cm} \small (Table 4 continued)}
\begin{center}
\begin{footnotesize}
\begin{tabular}
{|c|c|c|c|c|c|c|c|c|c|}  \hline
 ${s}$ & ${S}$ & ${T}$ & $\begin{array}{c} [5\, 1 ]
\otimes \\ \noalign{\vskip -2mm} [4\, 2] \\ \end{array} $ &
$\begin{array}{c} [4\,1\, 1] \\ \noalign{\vskip -2mm} \otimes [4
\, 2] \\ \end{array} $ & $\begin{array}{c} [3\, 3 ] \\
\noalign{\vskip -2mm} \otimes [6] \\ \end{array} $ &
$\begin{array}{c} [3\, 3 ] \\ \noalign{\vskip -2mm} \otimes [4 \,
2] \\ \end{array} $ & $\begin{array}{c} [3\, 2\, 1] \\
\noalign{\vskip -2mm} \otimes [5\, 1] \\ \end{array} $ &
$\begin{array}{c} [3\,2\, 1] \\ \noalign{\vskip -2mm} \otimes [4
\, 2] \\ \end{array} $ &
$\begin{array}{c}[2\,2\,1\,1 ] \\ \noalign{\vskip -2mm} \otimes [4\, 2] \\
\end{array} $ \\   \hline
\raisebox{-24.50ex}[0cm][0cm]{$-2$}& 0& 0& 1& 1& 0& 1& 2& 1& 1  \\
\cline{2-10}
 & 0 & 1 & 4 & 4 & 1 & 4 & 2 & 4 & 4  \\
\cline{2-10}
 & 0 & 2 & 1 & 1 & 0 & 1 & 1 & 1 & 1  \\
\cline{2-10}
 & 1 & 0 & 4 & 4 & 1 & 4 & 2 & 4 & 4  \\
\cline{2-10}
 & 1 & 1 & 6 & 6 & 1 & 6 & 5 & 6 & 6  \\
\cline{2-10}
 & 1 & 2 & 4 & 4 & 1 & 4 & 2 & 4 & 4  \\
\cline{2-10}
 & 2 & 0 & 2 & 2 & 0 & 2 & 1 & 2 & 2  \\
\cline{2-10}
 & 2 & 1 & 5 & 5 & 1 & 5 & 3 & 5 & 5  \\
\cline{2-10}
 & 2 & 2 & 2 & 2 & 1 & 2 & 2 & 2 & 2  \\
\cline{2-10}
 & 3 & 0 & 1 & 1 & 0 & 1 & 0 & 1 & 1  \\
\cline{2-10}
 & 3 & 1 & 1 & 1 & 0 & 1 & 1 & 1 & 1  \\
\cline{2-10}
 & 3 & 2 & 1 & 1 & 1 & 1 & 1 & 1 & 1  \\
\hline \raisebox{-15.50ex}[0cm][0cm]{$-3$}& 0& 1/2& 3& 3& 0&
3& 2& 3& 3 \\
\cline{2-10}
 & 0 & 3/2 & 2 & 2 & 1 & 2 & 1 & 2 & 2  \\
\cline{2-10}
 & 1 & 1/2 & 6 & 6 & 1 & 6 & 4 & 6 & 6  \\
\cline{2-10}
 & 1 & 3/2 & 4 & 4 & 1 & 4 & 3 & 4 & 4  \\
\cline{2-10}
 & 2 & 1/2 & 4 & 4 & 1 & 4 & 3 & 4 & 4  \\
\cline{2-10}
 & 2 & 3/2 & 3 & 3 & 1 & 3 & 3 & 3 & 3  \\
\cline{2-10}
 & 3 & 1/2 & 1 & 1 & 0 & 1 & 1 & 1 & 1  \\
\cline{2-10}
 & 3 & 3/2 & 1 & 1 & 1 & 1 & 1 & 1 & 1  \\
\hline \raisebox{-15.50ex}[0cm][0cm]{$-4$}& 0& 0& 2& 2& 0& 2&
0& 2& 2  \\
\cline{2-10}
 & 0 & 1 & 1 & 1 & 0 & 1 & 1 & 1 & 1   \\
\cline{2-10}
 & 1 & 0 & 2 & 2 & 0 & 2 & 2 & 2 & 2   \\
\cline{2-10}
 & 1 & 1 & 4 & 4 & 1 & 4 & 2 & 4 & 4   \\
\cline{2-10}
 & 2 & 0 & 2 & 2 & 1 & 2 & 2 & 2 & 2   \\
\cline{2-10}
 & 2 & 1 & 2 & 2 & 1 & 2 & 3 & 2 & 2   \\
\cline{2-10}
 & 3 & 0 & 0 & 0 & 0 & 0 & 1 & 0 & 0   \\
\cline{2-10}
 & 3 & 1 & 1 & 1 & 1 & 1 & 1 & 1 & 1   \\
\hline
\end{tabular}
\end{footnotesize}
\end{center}
%\label{tab1}
\end{table}

\begin{table}[htbp]
{\hspace*{1cm} \small (Table 4 continued)}
\begin{center}
\begin{footnotesize}
\begin{tabular}{|c|c|c|c|c|c|c|c|c|c|}
\hline ${s}$ & ${S}$ & ${T}$ & $\begin{array}{c} [5\, 1] \otimes
\\ \noalign{\vskip -2mm} [4\, 2] \\ \end{array} $ & $\begin{array}{c}
[4\,1\, 1] \\ \noalign{\vskip -2mm} \otimes [4\, 2] \\
\end{array} $
& $\begin{array}{c} [3\, 3] \\
\noalign{\vskip -2mm} \otimes [6] \\ \end{array} $ &
$\begin{array}{c} [3\, 3] \\ \noalign{\vskip -2mm} \otimes [4 \,
2] \\ \end{array} $ & $\begin{array}{c} [3\, 2\, 1] \\
\noalign{\vskip -2mm} \otimes [5\, 1] \\ \end{array} $ &
$\begin{array}{c} [3\,2\, 1] \\ \noalign{\vskip -2mm} \otimes [4
\, 2] \\ \end{array} $ &
$\begin{array}{c} [2\,2\,1\,1] \\ \noalign{\vskip -2mm} \otimes [4\, 2] \\
\end{array} $ \\   \hline
\raisebox{-6.50ex}[0cm][0cm]{$-5$}& 0& 1/2& 1& 1& 0& 1& 0& 1& 1  \\
\cline{2-10}
 & 1 & 1/2 & 1 & 1 & 0 & 2 & 1 & 2 & 2  \\
\cline{2-10}
 & 2 & 1/2 & 2 & 1 & 1 & 1 & 2 & 1 & 1  \\
\cline{2-10}
 & 3 & 1/2 & 0 & 0 & 1 & 0 & 1 & 0 & 0  \\
\hline \raisebox{-6.50ex}[0cm][0cm]{$-6$}& 0& 0& 0& 0& 0& 0&
0& 0& 0 \\
\cline{2-10}
 & 1 & 0 & 1 & 1 & 0 & 1 & 0 & 1 & 1  \\
\cline{2-10}
 & 2 & 0 & 0 & 0 & 0 & 0 & 1 & 0 & 0  \\
\cline{2-10}
 & 3 & 0 & 0 & 0 & 1 & 0 & 0 & 0 & 0  \\
\hline
\end{tabular}
\end{footnotesize}
\end{center}
%\label{tab4}
\end{table}

\vfill

\newpage

\begin{table}[htbp]
\begin{center}
\caption{\small {Quantum numbers of the candidates of low-lying
dibaryons with $L=0$ and $L=1$}} \vspace*{5mm}

\begin{tabular}{|c|c|c|c|c|c|c|} \hline
 $s$ & $S$  & $T$ & $L$ & $\pi$ & $J$ & $[f]_O\otimes[f]_{FS}$ \\
\hline 0 & 1 & 0 & 0 & $+$ & 1 & $[6 ]\otimes [3\, 3],\; [4\, 2]
\otimes[5\, 1],\; [4\, 2]\otimes [3\, 3], \; [ 2\, 2\, 2] \otimes
[3\, 3] $   \\ \hline
 0 & 3 & 0 & 0 & $+$ & 3 & $[6 ]\otimes [3\, 3],\; [4\, 2] \otimes
 [3\, 3], \; [ 2\, 2\, 2 ] \otimes [3\, 3] $   \\
\hline
 0 & 0 & 1 & 0 & $+$ & 0 & $ [6 ]\otimes [3\, 3],\; [4\, 2 ]
\otimes[5\, 1],\; [4\, 2 ]\otimes[3\, 3], \; [ 2\, 2\, 2 ] \otimes
[3\, 3] $   \\ \hline
 $-5$ & 0 & $\frac{1}{2}$ & 0 & $+$ & 0 & $[6 ] \otimes [3\, 3],\;
[4\, 2 ]\otimes[3\, 3], \; [ 2\, 2\, 2 ] \otimes [3\, 3] $   \\
\hline $-5$ & 1 &  $\frac{1}{2}$ & 0 & $+$ & 1 & $[6 ]\otimes [3\,
3],\; [4\, 2 ]\otimes[5\, 1],\; [4\, 2]\otimes[3\, 3], \; [ 2\,
2\, 2]\otimes [3\, 3]$   \\ \hline
 $-6$ & 0 & 0 & 0 & $+$ & 0 & $[6 ] \otimes [3\, 3],\; [4\, 2 ]
\otimes[3\, 3], \; [ 2\, 2\, 2 ]\otimes [3\, 3]$   \\  \hline $-2$
& 3 & 0 & 1 & $-$ & 2, 3, 4 & $[5\, 1 ] \otimes [4\,2],\; [4\, 1\,
1] \otimes [4\,2],\; [3\, 3] \otimes [4\,2], $                          \\
& & & & & & $[3\, 2\, 1]\otimes[4\,2],\; [2\, 2\, 1\, 1]\otimes[4\,2]$  \\
\hline $-6$ & 1 & 0 & 1  & $-$    & 0,\ 1,\ 2 & $[5\, 1]
\otimes[4\,2],\; [4\, 1\, 1]\otimes[4\,2],\; [3\, 3]\otimes
[4\,2],$                                                       \\
& & & & & & $[3\, 2\, 1 ]\otimes[4\,2],\; [2\, 2\, 1\, 1]
\otimes [4\,2]$                                                      \\
\hline
\end{tabular}
\end{center}
\end{table}

\end{document}